\documentclass[prl,twocolumn,a4paper,superscriptaddress,floatfix]{revtex4-1}

\pdfoutput=1

\usepackage{dcolumn,amsmath,xspace}
\PassOptionsToPackage{caption=false}{subfig}
\usepackage{subfig}
\usepackage{graphicx}
\usepackage{multirow}

\usepackage[utf8]{inputenc}

\usepackage{epstopdf}

\begin{document}
\title{Thickness-Dependent Polarization of Strained BiFeO$_\mathrm{3}$ Films with Constant Tetragonality}

\author{J. E. Rault}
\affiliation{CEA, DSM/IRAMIS/SPCSI, F-91191 Gif-sur-Yvette Cedex, France}
\author{W. Ren}
\affiliation{Physics Department and Institute for Nanoscience and Engineering, University of Arkansas, Fayetteville, Arkansas 72701, USA}
\affiliation{Department of Physics, Shanghai University, 99 Shangda Road, Shanghai 200444, China}
\author{S. Prosandeev}
\affiliation{Physics Department and Institute for Nanoscience and Engineering, University of Arkansas, Fayetteville, Arkansas 72701, USA}
\author{S. Lisenkov}
\affiliation{Department of Physics, University of South Florida, 4202 East Fowler Avenue, Tampa, FL 33620-5700, USA}
\author{D. Sando}
\affiliation{Unité Mixte de Physique CNRS/Thales, 1 Avenue Augustin Fresnel, Campus de l'Ecole Polytechnique, 91767 Palaiseau, France and Université Paris-Sud, 91405 Orsay, France}
\author{S. Fusil}
\affiliation{Unité Mixte de Physique CNRS/Thales, 1 Avenue Augustin Fresnel, Campus de l'Ecole Polytechnique, 91767 Palaiseau, France and Université Paris-Sud, 91405 Orsay, France}
\affiliation{Université d'Evry-Val d'Essonne, Boulevard François Mitterrand, 91025 Evry cedex, France}
\author{M. Bibes}
\affiliation{Unité Mixte de Physique CNRS/Thales, 1 Avenue Augustin Fresnel, Campus de l'Ecole Polytechnique, 91767 Palaiseau, France and Université Paris-Sud, 91405 Orsay, France}
\author{A. Barthélémy}
\affiliation{Unité Mixte de Physique CNRS/Thales, 1 Avenue Augustin Fresnel, Campus de l'Ecole Polytechnique, 91767 Palaiseau, France and Université Paris-Sud, 91405 Orsay, France}
\author{L. Bellaiche}
\affiliation{Physics Department and Institute for Nanoscience and Engineering, University of Arkansas, Fayetteville, Arkansas 72701, USA}
\author{N. Barrett}
\affiliation{CEA, DSM/IRAMIS/SPCSI, F-91191 Gif-sur-Yvette Cedex, France}

\begin{abstract}
We measure the ferroelectric polarization of BiFeO$_\mathrm{3}$ films down to 3.6~nm using low energy electron and photoelectron emission microscopy. The measured polarization decays strongly below a critical thickness of 5-7~nm predicted by continuous medium theory whereas the tetragonal distortion does not change. We resolve this apparent contradiction using first-principles-based effective Hamiltonian calculations. In ultrathin films the energetics of near open-circuit electrical boundary conditions, i.e.\ unscreened depolarizing field, drive the system through a phase transition from single out-of-plane polarization to nanoscale stripe domains. It gives rise to an average polarization close to zero as measured by the electron microscopy whilst maintaining the relatively large tetragonal distortion imposed by the nonzero polarization state of each individual domain.
\end{abstract}
\vspace*{4ex}

\pacs{77.80.-e  68.37.Xy  77.55.fp}
\keywords{Ferroelectricity, Bismuth Ferrite, PhotoEmission Microscopy, Low Energy Electron Microscopy, PiezoResponse Force Microscopy}
\maketitle
\newpage
A major issue for prospective nanoscale, strain-engineered~\cite{pan_enhancement_2004} ferroelectric devices~\cite{bibes_nanoferronics_2012} is the decrease of the polarization P$_\mathrm{r}$ of ultrathin films. The depolarizing field arising from uncompensated surface charges reduces or even suppresses ferroelectricity below a critical thickness~\cite{gerra_ionic_2006,junquera_critical_2003}. Ferroelectric capacitors for example may exhibit a critical thickness~\cite{kim_polarization_2005,petraru_wedgelike_2008}. Lichtensteiger \textit{et al.}~\cite{lichtensteiger_ferroelectricity_2005} have shown that the decrease in P$_\mathrm{r}$ in PbTiO$_\mathrm{3}$ (PTO) thin films between 20 and 2.4~nm on Nb-doped SrTiO$_\mathrm{3}$ (STO) substrates is concomitant with that of the tetragonality (ratio of the out-of-plane to in-plane lattice parameter c/a)). On La$_\mathrm{0.67}$Sr$_\mathrm{0.33}$MnO$_\mathrm{3}$ (LSMO), PTO polydomains were formed below 10~nm with high tetragonality~\cite{lichtensteiger_monodomain_2007}. The formation of a polydomain state has been suggested for SrRuO$_\mathrm{3}$/Pb(Zr,Ti)O$_\mathrm{3}$/SrRuO$_\mathrm{3}$ capacitors with Pb(Zr,Ti)O$_\mathrm{3}$ thicknesses below 15~nm~\cite{nagarajan_scaling_2006}. Pertsev and Kohlstedt showed the importance of misfit strain for the critical thickness of the monodomain-polydomain stability for PTO and BaTiO$_\mathrm{3}$~\cite{pertsev_elastic_2007}. Using piezoresponse force microscopy (PFM), BiFeO$_\mathrm{3}$ (BFO) films have been shown to remain ferroelectric down to a few unit cells~\cite{bea_ferroelectricity_2006,chu_ferroelectric_2007,maksymovych_ultrathin_2012} with both the polarization and the slope of the piezoresponse hysteresis loop scaling with tetragonality. However, PFM is very local and can only provide indirect, semiquantitative estimates of the polarization. Imperfect tip surface contact can contribute to polarization suppression via the depolarizing field. Direct electrical measurements of the polarization-field (P(E)) loop in ultrathin ferroelectric films are a challenge because of leakage current for thicknesses below a few tens of~nm~\cite{bea_ferroelectricity_2006,kim_effect_2008}. They become impossible in the tunneling regime for ultra-thin films (5~nm or less) which, furthermore, is of the same order as the critical thickness, h$_\mathrm{eff}$, estimated from Landau-Ginzburg-Devonshire (LGD) elastic theory for polarization stability~\cite{maksymovych_ultrathin_2012,bratkovsky_abrupt_2000}. BFO can accommodate in-plane compressive strain via out-of-plane extension and through oxygen octahedron rotation about $\langle 111 \rangle$~\cite{infante_bridging_2010}, a degree of freedom not available in P4mm PTO films. This interplay between strain, tetragonality and octahedra rotations leads to an unexpected decrease of T$_\mathrm{C}$ with strain, at odds with the variation of c/a ratio. Thus the relationship between structural parameters and the remnant out-of-plane polarization in very thin films remains an open question. 

\begin{figure}
  \centering
  	\subfloat{\label{fig:XRD_exp}\includegraphics[scale=0.378]{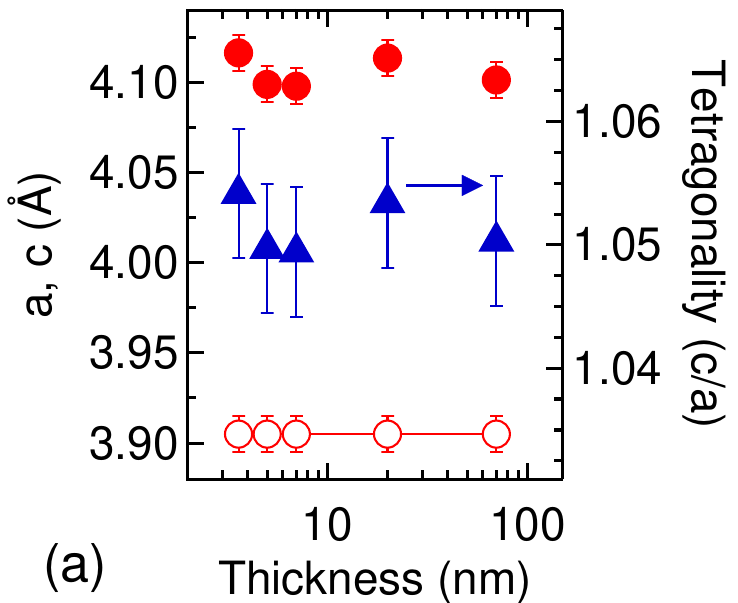}}
  	\subfloat{\label{fig:XPS}\includegraphics[scale=0.378]{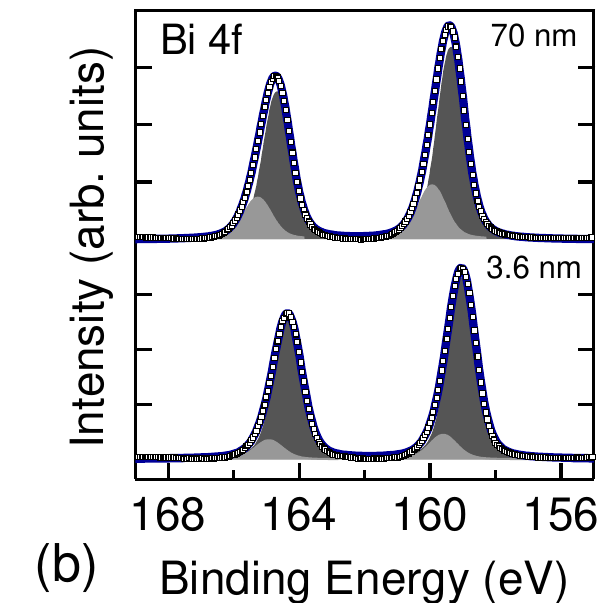}} \quad
  	\subfloat{\label{fig:PEloop}\includegraphics[scale=0.378]{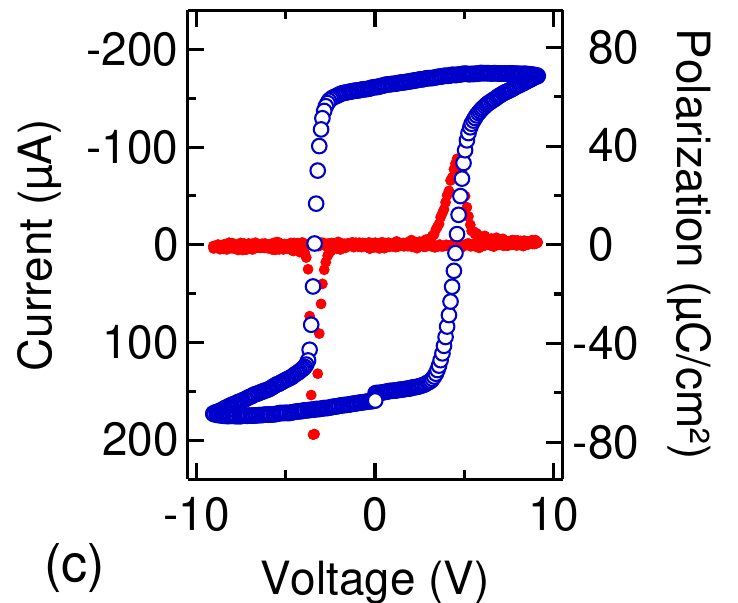}}
  	\subfloat{\label{fig:piezoLoop}\includegraphics[scale=0.378]{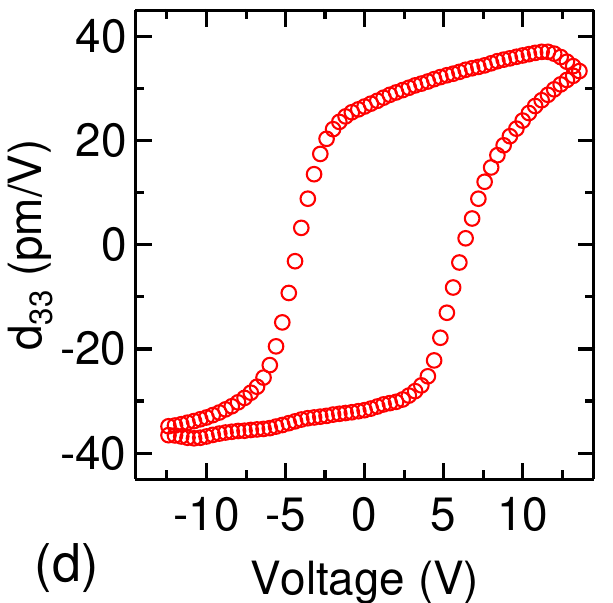}}
  \caption{(a) Evolution of pseudocubic lattice parameters a and c and c/a ratio with thickness, (b) Bi~4f spectra for 70 and 3.6~nm films showing two components (surface in light grey, bulk in dark grey) for each spin-orbit core-level, (c) Polarization-voltage and current-voltage hysteresis loop of BFO(70~nm)/LSMO(20~nm)//STO(001) (d) Piezoresponse hysteresis loop (local measurement under the PFM tip).}
  \label{fig:CARAC}
\end{figure}

In this Letter, we have studied the polarization of BFO films from 70 to 3.6~nm thick using a combination of x-ray diffraction (XRD), mirror electron microscopy (MEM) and photoelectron emission microscopy (PEEM). The electron microscopy techniques provide full-field imaging of the electrostatic potential above the surface and the work function whereas the tetragonality is measured by XRD. The results are interpreted in the light of a three-dimensional (3D) generalization of previously developed dead layer model for thin films within the framework of continuous medium theory that predicts a fast decrease of the polarization when decreasing the thickness. Interestingly, the extremely low polarization below h$_\mathrm{eff}$ does not scale with the tetragonality and is explained using first principles-based effective Hamiltonian calculations which show that as a function of screening the films undergo a phase transition from single to nanoscale stripe domains with an overall out-of-plane polarization close to zero.\\
Bilayers of BFO/LSMO were epitaxially grown on (001)-oriented STO substrates by pulsed laser deposition using a frequency tripled (h$\nu$ = 355~nm) Nd-doped yttrium aluminium garnet (Nd:YAG) laser at a frequency of 2.5~Hz~\cite{bea_ferroelectricity_2006}. The 20~nm thick LSMO layer serves as a metallic bottom electrode for ferroelectric characterization. XRD measurements on 70~nm to 3.6~nm-thick thin films were performed to track the out-of-plane parameter and c/a ratio (Fig.~\ref{fig:XRD_exp}). The c/a increases slightly from 1.050 for the 70~nm film to 1.053 for 7~nm, then remains constant down to 3.6~nm. This contrasts dramatically with the behavior of PTO reported in~\cite{lichtensteiger_ferroelectricity_2005} where c/a decreases with thickness. 
The chemistry of the films was measured by X-ray Photoelectron Spectroscopy (XPS). Figure~\ref{fig:XPS} shows spectra from Bi~4f core-level for thickest (70~nm) and thinnest (3.6~nm) films. The spectra are virtually identical for both films (for intermediate thicknesses, see \footnote{See Supplemental Material at [URL].}) showing that the chemical state and stoichiometry do not change. Bi~4f spectra have a thickness \emph{independent} component shifted by 0.6~eV to higher binding energy, suggesting that our strained thin films do not exhibit the several nanometer thick skin observed on single crystals~\cite{marti_skin_2011}. C~1s spectra show that contamination of the BFO surface is similar for every thickness suggesting a similar contribution to extrinsic screening in all films~\cite{Note1}.

\begin{figure}
  \centering
  \subfloat{\label{fig:PFM}\includegraphics[scale=0.292]{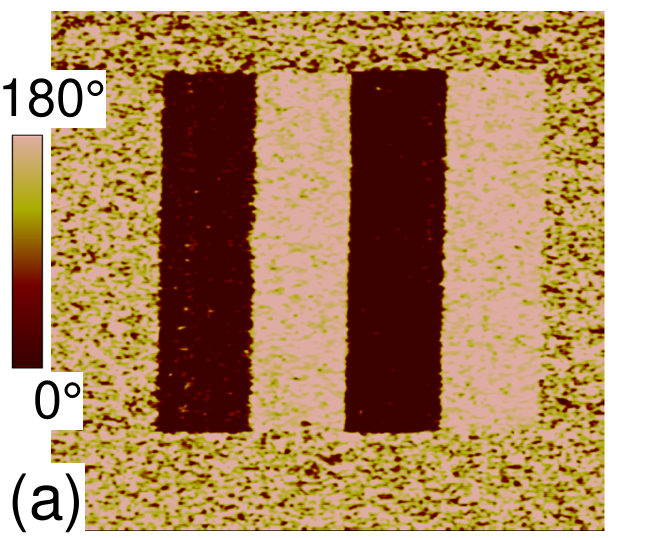}}\hfill
  \subfloat{\label{fig:LEEM_image}\includegraphics[scale=0.292]{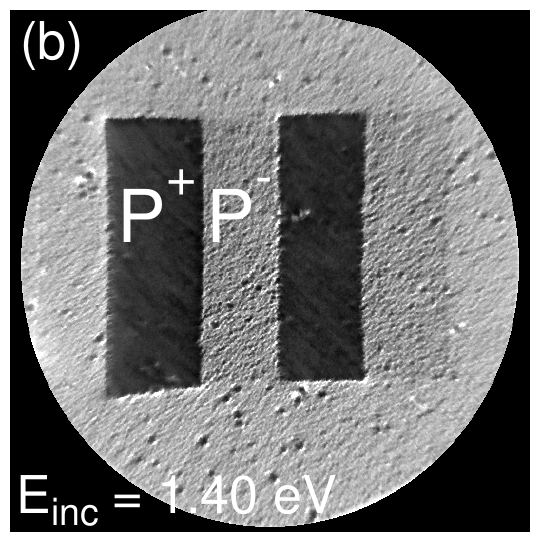}}\hfill
  \subfloat{\label{fig:PEEM_image}\includegraphics[scale=0.292]{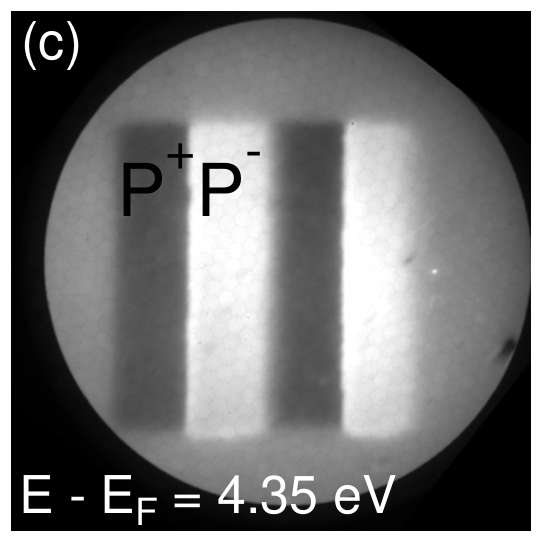}}
  \caption{(a) PFM phase image of written P$^\mathrm{+}$/P$^\mathrm{-}$ domains. Each domain is 5x20~$\mathrm{\mu m^2}$, (b) LEEM image for E$_{\mathrm{inc}}$ = 1.40~eV, (c) PEEM image at E - E$_\mathrm{F}$ = 4.35~eV.}
  \label{fig:PFM_LEEM_PEM}
\end{figure}

For the 70~nm BFO film, the ferroelectric properties were investigated by standard polarization versus electric field P(E) loops (Fig.~\ref{fig:PEloop}). The piezo-response hysteresis loops are shown in Fig.~\ref{fig:piezoLoop}. They are position independent and exhibit similar coercive values as nonlocal P(E) loops, attesting to sample homogeneity. In a (001) BFO film P$^\mathrm{+}$ and P$^\mathrm{-}$ states are the projections of $\langle 111 \rangle$ polarization along [001]. Poling of micron sized domains was performed by applying a d.c.~voltage higher than the coercive voltage (inferred from the piezoresponse loops) on the tip while the bottom electrode was grounded. PFM imaging was carried out at an excitation frequency of 4-7~kHz and an a.c.~voltage of 1~V. No morphology change occurred during poling as checked by Atomic Force Microscopy.
A low-energy electron microscope (LEEM, Elmitec GmbH) was used to measure the electron kinetic energy of the MEM (reflected electrons)-LEEM (backscattered electrons) transition with a spatial resolution of 30~nm. The transition energy is a measure of electrostatic potential just above the sample surface~\cite{cherifi_imaging_2010} and depends on polarization and the screening of polarization-induced surface-charge~\cite{krug_extrinsic_2010}. It allows a noncontact estimation of the out-of-plane polarization for tunneling films, otherwise inaccessible to standard electrical methods. All experiments were done at least two days after domain writing to ensure that the observed contrast is not due to residual injected charges. 

\begin{figure}
  \centering
  \subfloat{\label{fig:MEM_LEEM}\includegraphics[scale=0.31]{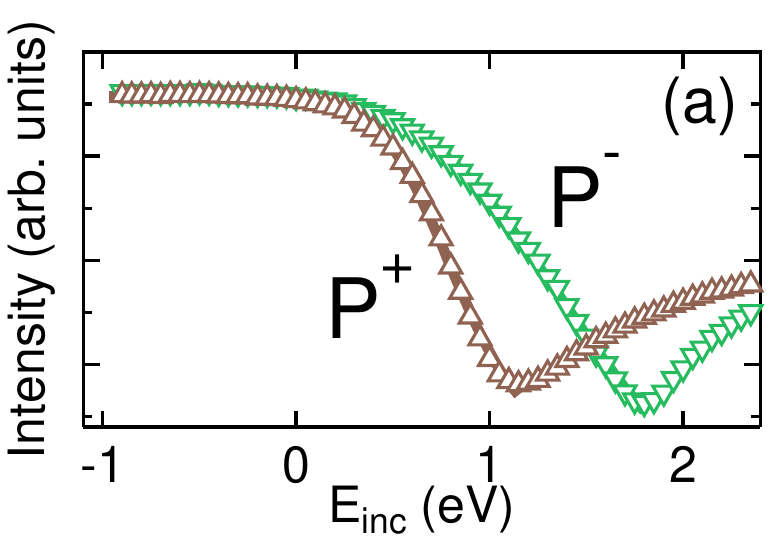}}\hfill
  \subfloat{\label{fig:SV_MAP}\includegraphics[scale=0.31]{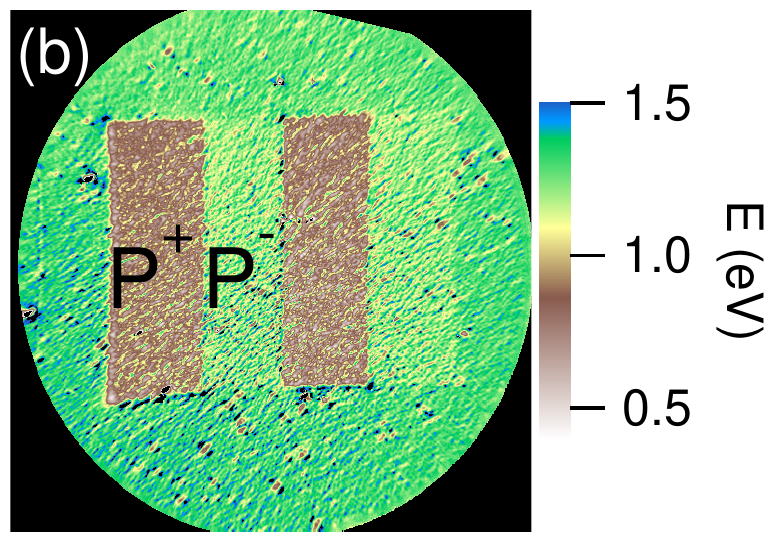}}\hfill
  \caption{(a) Reflectivity spectra extracted from the P$^\mathrm{+}$ and P$^\mathrm{-}$ domains, (b) MEM-LEEM transition map obtained from the image series (70~nm thin film).}
  \label{fig:LEEM}
\end{figure}

Figure~\ref{fig:LEEM_image} shows a typical LEEM image with a field of view (FoV) of 33~$\mathrm{\mu m}$ for incident electron energy (E$_{\mathrm{inc}}$) of 1.40~eV. The observed contrast reproduces well the PFM image of Fig.~\ref{fig:PFM}. A full image series across the MEM-LEEM transition (E) was acquired by varying E$_{\mathrm{inc}}$ from -2.0 to 3.0~eV. Figure~\ref{fig:MEM_LEEM} displays the electron reflectivity curves showing the MEM (high reflectivity) to LEEM (low reflectivity) transition for the P$^\mathrm{+}$ (brown upwards triangles, E = 0.75~eV) and P$^\mathrm{-}$ (green downwards triangles, E = 1.20~eV) domains. Using complementary error function fits we obtain MEM-LEEM transition maps showing clear contrast in the electrostatic potential above the surface between the P$^\mathrm{+}$, P$^\mathrm{-}$ and unwritten regions (Fig.~\ref{fig:SV_MAP}).

\begin{figure}
  \centering
  \subfloat{\label{fig:Thresh}\includegraphics[scale=0.31]{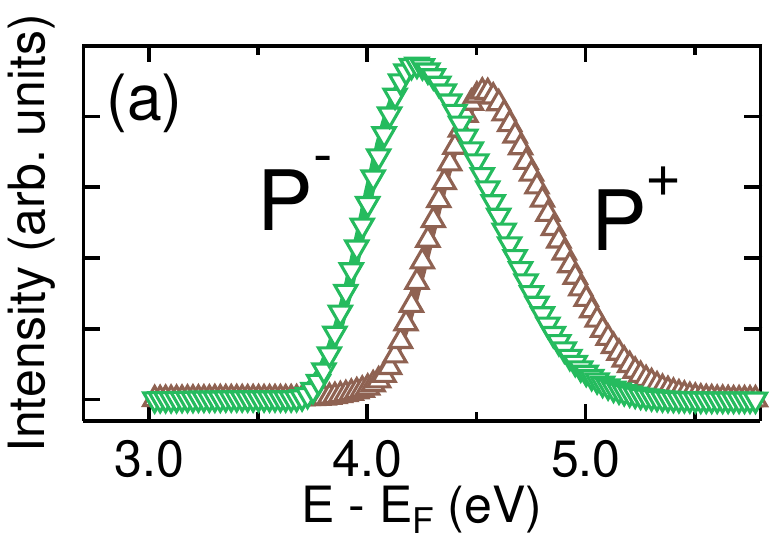}}\hfill
  \subfloat{\label{fig:WF_MAP}\includegraphics[scale=0.31]{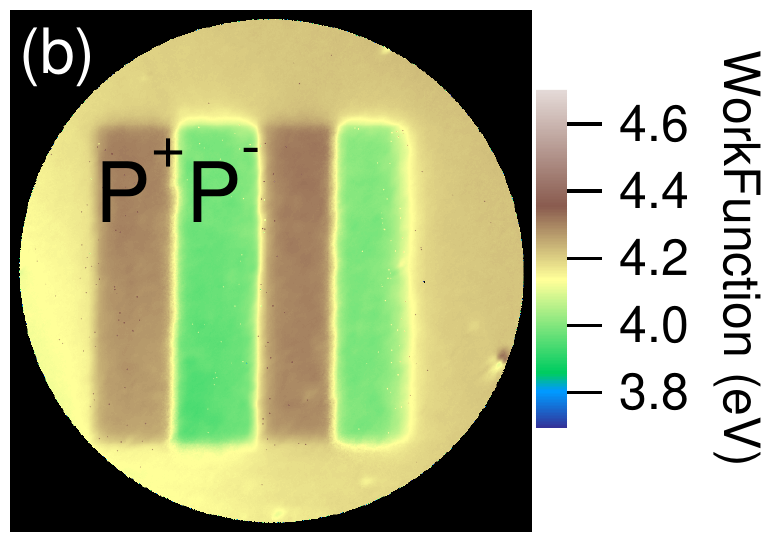}}\hfill
  \caption{(a) Threshold spectra extracted from P$^\mathrm{+}$ and P$^\mathrm{-}$ domains (b) Work function map obtained from the threshold image series (70~nm thin film).}
  \label{fig:PEEM}
\end{figure}

The energy filtered PEEM experiments used a NanoESCA X-PEEM (Omicron Nanotechnology GmbH). PEEM of the photoemission threshold gives a direct, accurate ($\pm$ 20~meV) and nondestructive map of the work function~\cite{mathieu_microscopic_2011} that may depend, for example, on domain polarization~\cite{barrett_influence_2010}. Image series were acquired over the photoemission threshold region with a mercury lamp excitation ($h\nu$ = 4.9~eV). The lateral resolution was estimated to be 200~nm and energy resolution 200~meV. Figure~\ref{fig:PEEM_image} shows a typical PEEM image of the prepoled P$^\mathrm{+}$ and P$^\mathrm{-}$ regions for the 70~nm BFO film. The energy contrast between oppositely polarized domains fits the PFM image except at the domain boundary where the lateral electric field induced by a P$^\mathrm{+}$/P$^\mathrm{-}$ domain wall deflects electrons~\cite{nepijko_peculiarities_2001}. We extract the threshold from the pixel-by-pixel spectra using a complementary error function to model the rising edge of the photoemission (Fig.~\ref{fig:Thresh}). Figure~\ref{fig:WF_MAP} maps the work function in the P$^\mathrm{+}$, P$^\mathrm{-}$ and as-grown regions. 

\begin{figure}
  \centering
  \subfloat{\label{fig:SV_WF}\includegraphics[scale=0.36]{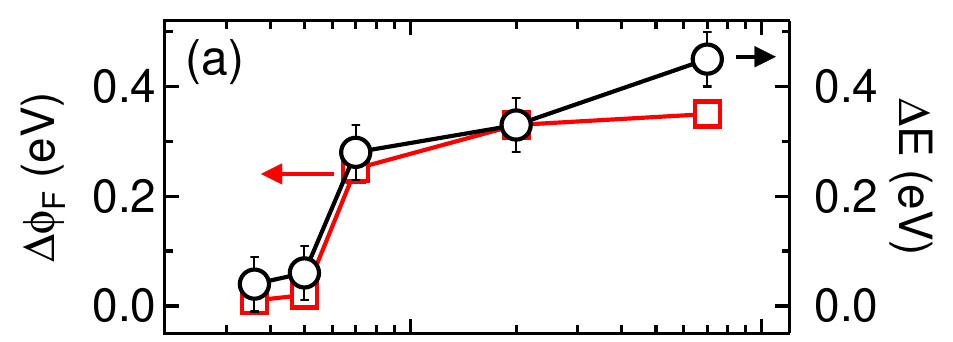}}\quad
  \subfloat{\label{fig:P_PMAX}\includegraphics[scale=0.36]{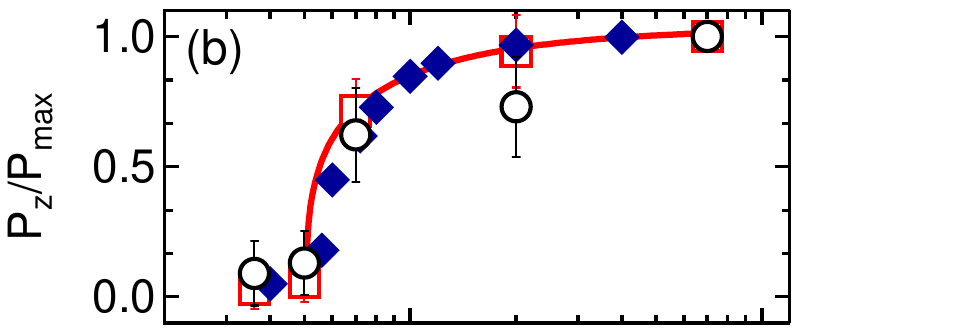}}\quad
  \subfloat{\label{fig:beta_t}\includegraphics[scale=0.36]{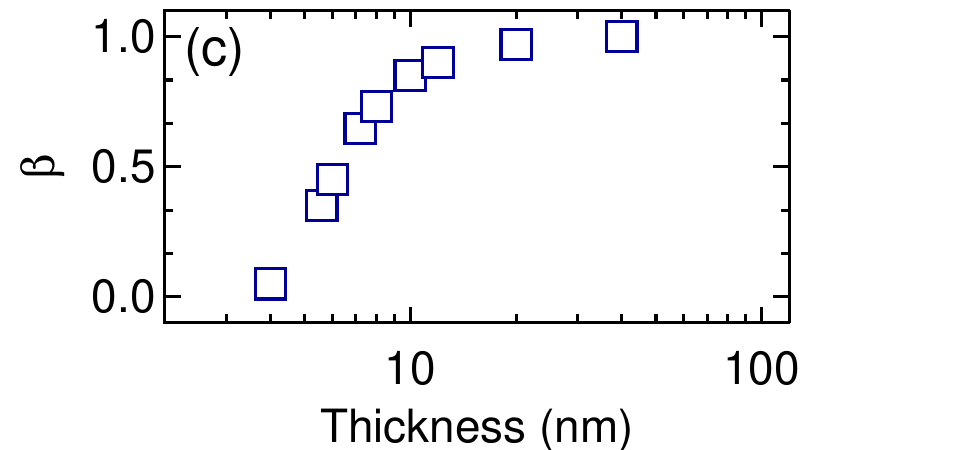}}
  \caption{Thickness dependence of (a) $\Delta\Phi_{F}$ (red squares) and $\Delta$E (black circles), (b) P$_\mathrm{z}$/P$_\mathrm{max}$ calculated from PEEM (red squares) and MEM-LEEM (black circles). Red curve is fit to PEEM/LEEM data with h$_\mathrm{eff}$ = 5.6~nm. Blue diamonds are P$_\mathrm{z}$/P$_\mathrm{max}$ values used for numerical simulations, (c) screening coefficient $\beta$ calculated from experimental P$_\mathrm{z}$/P$_\mathrm{max}$ values.}
  \label{fig:bilan}
\end{figure}

The difference in the MEM-LEEM transition of the P$^\mathrm{+}$ and P$^\mathrm{-}$ regions, $\Delta$E, varies from 450~meV for the 70~nm film to 25~meV for the 3.6~nm film and is plotted in Fig.~\ref{fig:SV_WF} (black circles, right axis). The mean work function difference measured in PEEM between P$^\mathrm{+}$ and P$^\mathrm{-}$ domains, $\Delta\Phi_{F} = \Phi_{F}(P^+) - \Phi_{F}(P^-)$, is plotted as a function of thickness in Fig.~\ref{fig:SV_WF} (left axis). While $\Delta\Phi_{F}$ is 300~meV between 70~nm and 7~nm, between 7 and 5~nm it drops to 20~meV. 

The polarization charges at the BFO surface are screened over a so-called dead layer leading to an inward ($P^+$) or outward ($P^-$) surface dipole. By measuring the work function (or surface potential) \emph{difference} between two opposite domains, our method allows a direct measurement of the polarization-induced dipoles since any averaged nonferroelectric contribution is canceled. The surface dipole difference, hence the surface potential or work function difference, is proportional to the difference in polarization charges when going from the $P^+$ to the $P^-$ domains:
\begin{equation} \label{eq:1}
\Delta\Phi_{F}, \Delta E \propto \frac{e}{\epsilon_0} \left(P^+.d^+ - P^-.d^- \right) \approx 2\frac{e}{\epsilon_0} P_r.d
\end{equation}
where $P^{+,-}$ and $d^{+,-}$ are the polarization and dead layer thickness for the upward, downward domains; $P_r$ is the average magnitude of the polarization in the two poled domains and d is the average dead layer thickness. For the sake of generality, one can take into account electronic screening via a high-frequency dielectric permittivity, but it would still leave a linear relation between polarization and $\Delta\Phi_{F}$, $\Delta$E. P$_\mathrm{z}$/P$_\mathrm{max}$, where P$_\mathrm{z}$ is the measured out-of-plane polarization and P$_\mathrm{max}$ the value for the 70~nm film, is plotted as a function of film thickness in Fig.~\ref{fig:P_PMAX}. By comparison with Fig.~\ref{fig:XRD_exp}, the drop of average polarization between 7 and 5~nm does not result from a decrease in the c/a ratio, contrary to PTO thin films~\cite{lichtensteiger_ferroelectricity_2005}. Here the c/a ratio increases for thinner films and is constant at 1.054 below 5~nm. If there were no polarization then it should be about 1.03. However, PTO is almost fully relaxed whereas BFO is compressively strained. Secondly, in BFO, the polarization deviates appreciably from the [001] direction and is the macroscopic average of four $\langle 111 \rangle$ type distortions. We have therefore generalized the 1D dead layer LGD model of Bratkovsky and Levanyuk~\cite{bratkovsky_abrupt_2000} to the 3D polarization case. It gives the following relation for thickness dependence of polarization~\cite{Note1}:
\begin{equation} \label{eq:2}
\frac{P_z}{P_{max}} = A \sqrt{B + \sqrt{1 - \frac{h_{eff}}{h}}}
\end{equation}
where h$_\mathrm{eff}$ is the effective thickness below which the macroscopic P$_\mathrm{z}$ goes to zero, and A, B are fitting parameters. A good fit to the data is obtained with h$_\mathrm{eff}$ = 5.6~nm (see Fig.~\ref{fig:P_PMAX}, red curve), compared with 2.4~nm for PTO. 

To understand why the measured polarization suddenly drops in ultrathin strained (001) BFO films, while the axial ratio is still very large, we have conducted first-principles-based, effective Hamiltonian calculations~\cite{prosandeev_kittel_2010,albrecht_ferromagnetism_2010,kornev_finite-temperature_2007} that take into account free surfaces~\cite{prosandeev_kittel_2010}. We used the lattice parameter of the STO substrate for the pseudo-cubic in-plane lattice constant of BFO, leading to a misfit strain of -1.8\%, in agreement with the experimental value. The calculation includes the local electric dipoles, the strain tensor and tilting of the oxygen octahedra. The electrical boundary conditions are governed by a coefficient denoted as $\beta$ described in Ref.~\onlinecite{ponomareva_atomistic_2005}. Practically, $\beta$ can vary between 0 (ideal open-circuit, maximal depolarizing field) and $\beta$ = 1 (ideal short-circuit, fully screened depolarizing field). To determine $\beta$ for each of our grown films we first extract the P$_\mathrm{z}$/P$_\mathrm{max}$ values from a B-spline interpolation of the experimental data (Fig.~\ref{fig:P_PMAX}, blue diamonds) and then vary $\beta$ in the calculations until the predicted P$_\mathrm{z}$/P$_\mathrm{max}$ perfectly agrees with the experimentally extracted one. Figure~\ref{fig:beta_t} shows the resulting $\beta$ values. $\beta$ decreases with thickness, indicating that the observed decrease of polarization is related to imperfect screening of the depolarizing field. The vanishing of the overall z-component of the polarization (which occurs experimentally for thicknesses lower than 5.6~nm, see Fig.~\ref{fig:P_PMAX}) is associated with values of $\beta$ lower than 0.4 (see Fig.~\ref{fig:beta_t}).

\begin{figure}
  \centering
  \subfloat{\label{fig:E_beta}\includegraphics[scale=0.38]{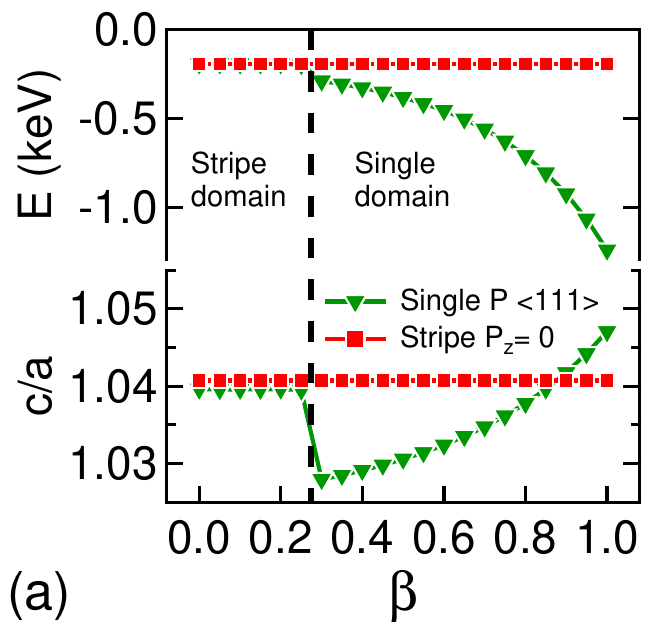}}\hfill
  \subfloat{\label{fig:P_beta}\includegraphics[scale=0.38]{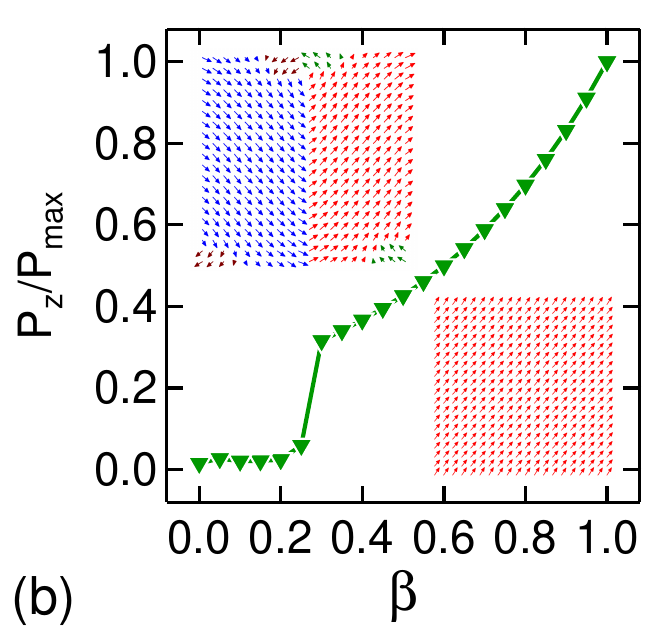}}\hfill
  \caption{(a) (top panel) Slab energy with $\beta$ showing a transition from single domain (out-of-plane polarization, green triangles) to stripe domains (no total out-of-plane polarization, red squares) below $\beta_{crit}$ = 0.30, (bottom panel) theoretical c/a ratio with $\beta$ for the two phases, (b) Evolution of P$_\mathrm{z}$/P$_\mathrm{max}$ with $\beta$, inset: domain morphology (upper left, stripe domains; bottom right, single domain). All these data are for a 20$\times$20$\times$20 slab, i.e.\ a film with a thickness of 8~nm.}
  \label{fig:theory}
\end{figure}

To understand what happens for these $\beta$ values, we performed additional first-principles-based effective Hamiltonian calculations on a single 20$\times$20$\times$20 supercell (i.e.\ with a thickness of 8~nm) allowing $\beta$ to vary. This supercell was chosen because around 8~nm the polarization is very sensitive to the thickness (Fig.~\ref{fig:P_PMAX}). The results are shown in Fig.~\ref{fig:theory}. At a critical value of $\beta$ of $0.275 \pm 0.025$ the BFO supercell goes from a phase with a uniform out-of-plane polarization to a stripe domain phase with a vanishing overall out-of-plane polarization. Fig.~\ref{fig:E_beta} displays the energy of these two phases as a function of $\beta$. The monodomain phase is energetically more favorable than the stripe nanodomains for $\beta$ above 0.30 and less for smaller $\beta$ values. The predicted evolution of the c/a ratio, and of the overall P$_\mathrm{z}$/P$_\mathrm{max}$, with $\beta$ for single and stripe domain phases are shown in Figs.~\ref{fig:E_beta} and~\ref{fig:P_beta}, respectively. Interestingly, a continuous ferroelectric to paraelectric transition would lead to a large monotonic decrease of tetragonality (Fig.~\ref{fig:E_beta}, green triangles), which we do not measure below h$_\mathrm{eff}$. Rather, the transformation from ferroelectric monodomains to nanostripe domains leads to a (large) c/a similar to the one associated with short-circuit-like conditions (i.e.\ for which $\beta$ is close to 1). Such results are consistent with our experimental findings that c/a does not vary between 70~nm and 3.6~nm, and explains that such insensitivity to strain is likely due to the formation of nanostripe domains. The single to stripe domain transition explains the loss of contrast in electronic microscopy observed in LEEM and PEEM contrast between 7 and 5~nm, because these regions do not possess any overall z-component of the polarization.
The stripes have a typical dimension of a few nanometers, which is below the lateral resolution of our experiments (The top left inset of Fig.~\ref{fig:P_beta} shows the morphology of these domains). However, stripe domains in BFO thin films close to the h$_\mathrm{eff}$ value have been observed by PFM~\cite{catalan_fractal_2008}. For such thin films, one might also ask to what extent the screening at the LSMO/BFO interface affects the measured polarization. Transmission electron microscopy of the interface between LSMO and a 3.2~nm BFO film suggests that the first three BFO unit cells are screened by the interface charge~\cite{chang_atomically_2011}. This also fits nicely with our experimental observation of an abrupt decrease in polarization starting at 7~nm, 1.4~nm above the calculated h$_\mathrm{eff}$.

In summary, we have measured the polarization in ultrathin strained (001) BFO films using PEEM and LEEM. The polarization drops abruptly below a critical thickness h$_\mathrm{crit}$ whereas the tetragonality has a high constant value. A first-principles-based effective Hamiltonian approach suggests that BFO exhibits a first order phase transition to stripe domains at h$_\mathrm{crit}$ = 5.6~nm, corresponding to a screening factor, $\beta$, below 0.35. This model fits the experimental measurement of the average polarization and the c/a ratio.

\begin{acknowledgments}
J.R. is funded by a CEA Ph.D. Grant CFR. This work was supported by the ANR projects "Meloïc", "Nomilops" and "Multidolls". W.R. acknowledges the Eastern Scholar Professorship at Shanghai Institutions of Higher Education, Shanghai Municipal Education Commission, and support from National Natural Science Foundation of China under Grant No. 11274222. L. B. thanks the financial support of ARO Grant No. W911NF-12-1-0085, and ONR Grants No. N00014-11-1-0384, N00014-12-1-1034 and No. N00014-08-1-091. We also acknowledge DOE, Office of Basic Energy Sciences, under Contract No. ER-46612 and NSF Grants No. DMR-1066158 and No. DMR-0701558, for discussions with scientists sponsored by these grants. Some computations were made possible thanks to the MRI Grant No. 0722625 (NSF), the ONR Grant No. N00014-07-1-0825 (DURIP) and a Challenge grant (DOD). We thank E. Jacquet, C. Carrétéro and H. Béa for assistance in sample preparation; K. Winkler, B. Krömker (Omicron Nanotechnology), C. Mathieu, D. Martinotti for help with the PEEM/LEEM experiments; and P. Jégou for the XPS measurements. 
\end{acknowledgments}

\bibliography{./biblio_BFO}

\clearpage

\onecolumngrid
\appendix

\graphicspath{{../img/supp_matt}}

\section*{Supplementary Materials}
\vspace*{4ex}

\subsection{X-Ray PhotoEmission Spectroscopy for Every Film Thickness}

\begin{figure}[h]
  \centering
  \subfloat[Bi 4f]{\label{fig:Bi4f}\includegraphics[scale=0.35]{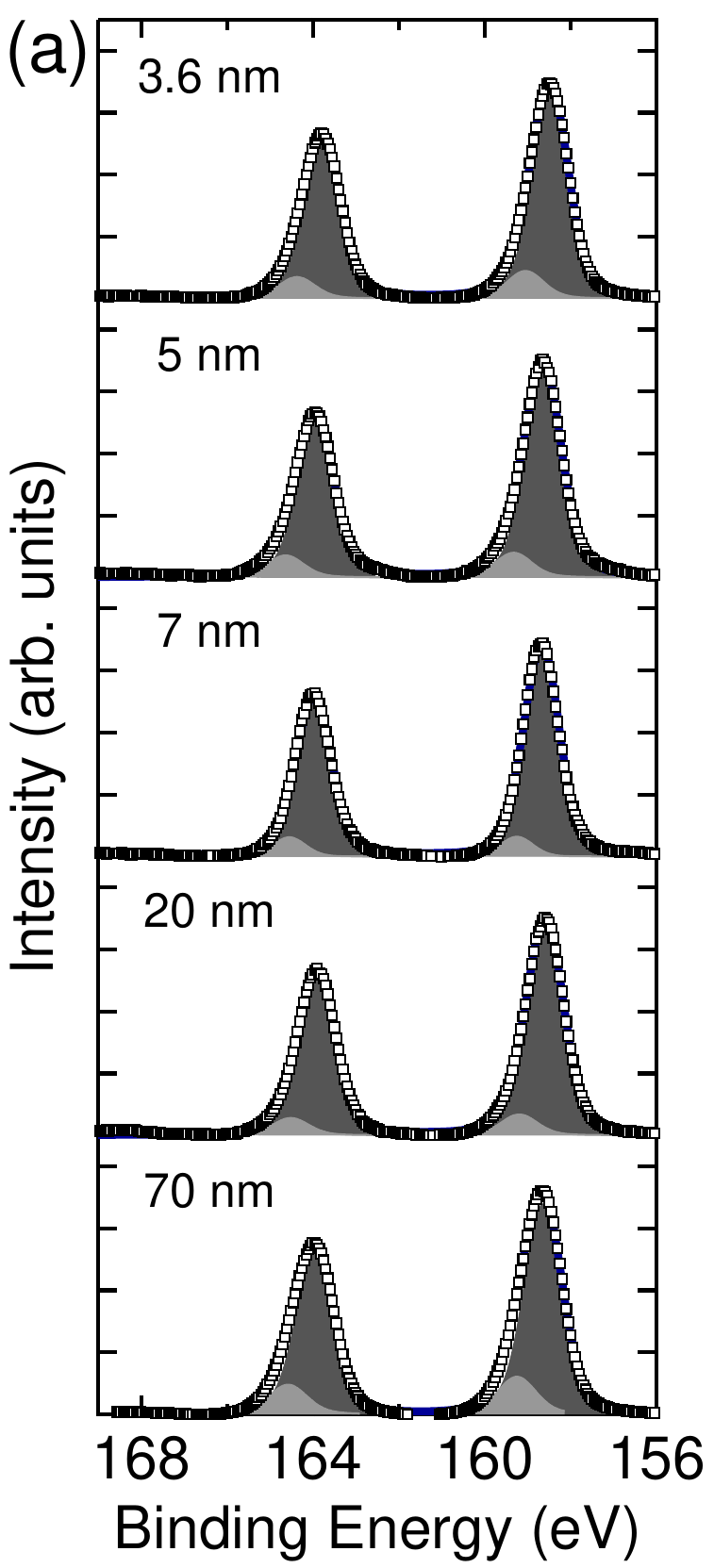}}\hfill
  \subfloat[Fe 2p]{\label{fig:Fe2p}\includegraphics[scale=0.35]{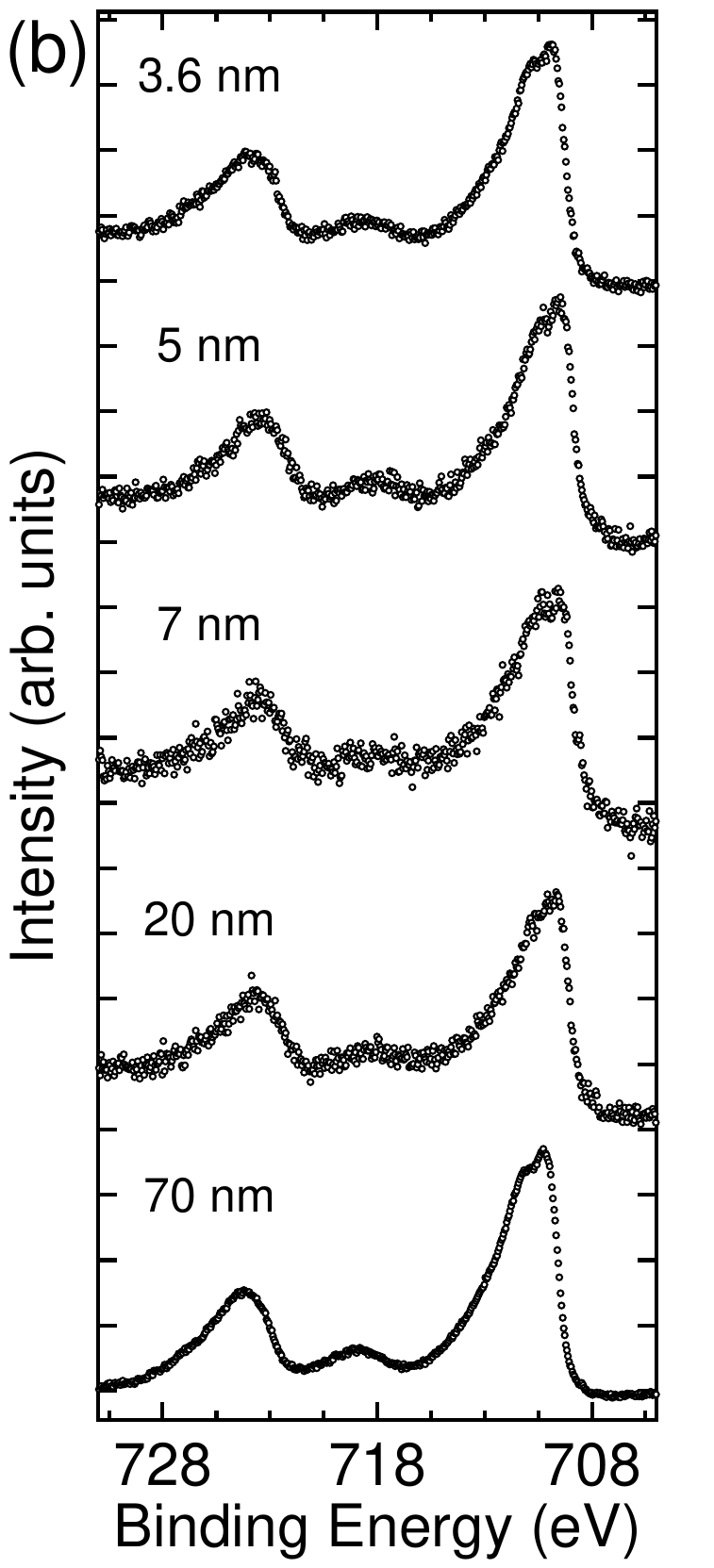}}\hfill
  \subfloat[C 1s]{\label{fig:C1s}\includegraphics[scale=0.35]{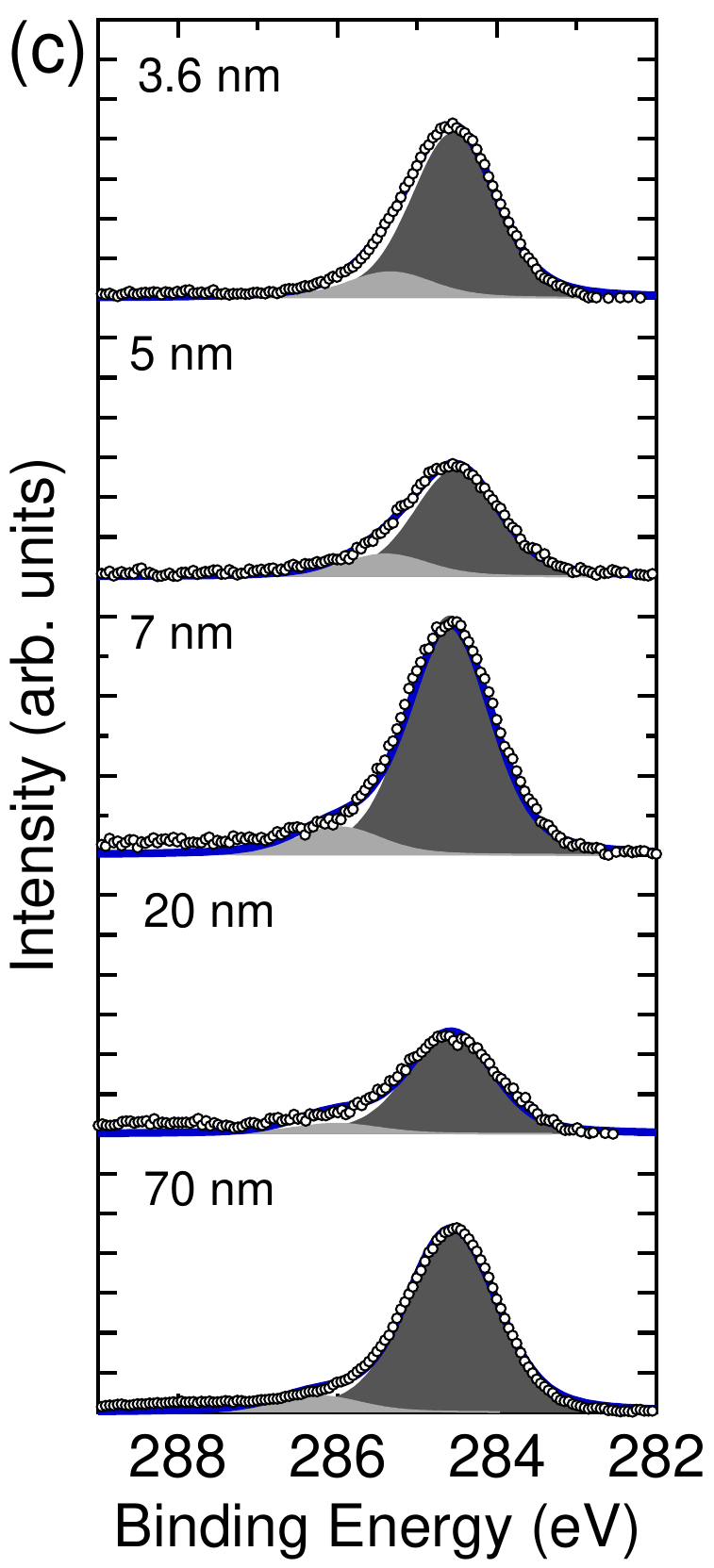}}\hfill
  \caption{XPS spectra of BiFeO$_{\mathrm{3}}$ thin films for every thickness.}
  \label{fig:XPS_total}
\end{figure}

X-ray PhotoEmission Spectroscopy (XPS) was carried out using a Kratos Ultra DLD with monochromatic Al K$\alpha$ (1486.7~eV). The analyzer pass energy of 20~eV gave an overall energy resolution (photons and spectrometer) of 0.35~eV. The sample is at floating potential and a charge compensation system was used. The binding energy scale was calibrated using a clean gold surface and the Au 4f$_{\mathrm{7/2}}$ line at 84.0~eV as a reference. A take-off angle of 90$^{\circ}$, i.e., normal emission, was used for all spectra presented. The XPS spectra show that the chemical environment was identical within 1\% (see~\ref{fig:Bi4f} and~\ref{fig:Fe2p}) for every film. Krug et al.~\cite{krug_extrinsic_2010} pointed out the importance of adsorbates on LEEM and PEEM measurements. Figure~\ref{fig:C1s} shows that surface contamination is similar in 3.6 (low contrast) and 70~nm (high contrast) thin films indicating strongly that the disappearance of ferroelectric contrast is not due to differential contamination. Moreover, the 5~nm film has the lowest carbonates concentration and still shows weak ferroelectric contrast in LEEM/PEEM experiments (see Table~\ref{tab:C1s}). 

\begin{table}[h]
	\centering
		\begin{tabular}{|c|c|}
			\hline
			\hspace{1cm} Thickness (nm) \hspace{1cm} & \hspace{1cm} $\frac{I_{C1s}}{\sigma_{C1s}} / \frac{I_{Bi4f}}{\sigma_{Bi4f}}$  \hspace{1cm} \\
			\hline
			3.6 & 2.22 \\
			5.0 & 1.49 \\
			7.0 & 3.30 \\
			20 & 1.40 \\
			70 & 2.20 \\
			\hline
		\end{tabular}
	\caption{C~1s to Bi~4f ratio calculated from XPS spectra}
	\label{tab:C1s}
\end{table}

\clearpage

\subsection{PFM/PEEM/LEEM Data for Every Film Thickness}

\begin{figure}[h]
  \centering
  \subfloat[70 nm]{\label{fig:PFM_70nm}\includegraphics[scale=0.3]{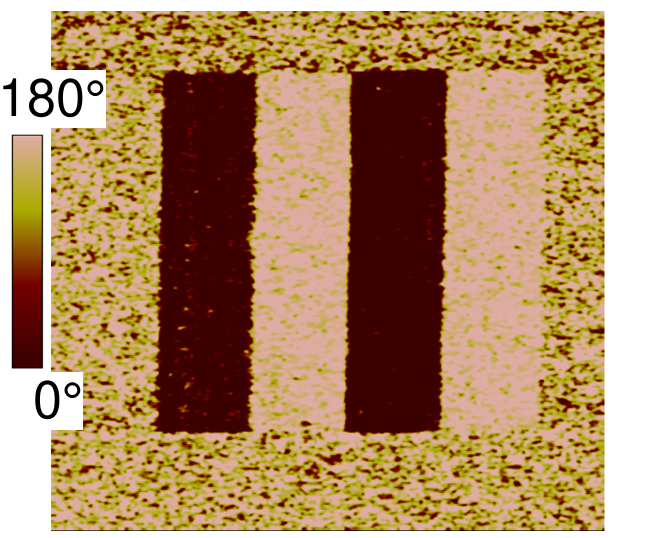}}\hfill
  \subfloat[20 nm]{\label{fig:PFM_20nm}\includegraphics[scale=0.3]{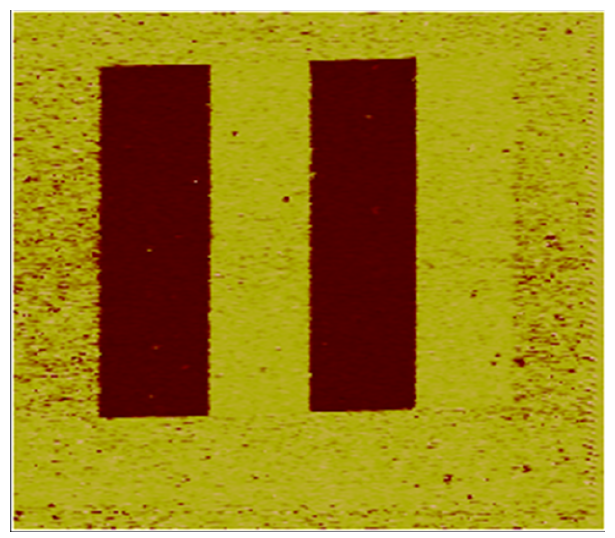}}\hfill
  \subfloat[7 nm]{\label{fig:PFM_7nm}\includegraphics[scale=0.3]{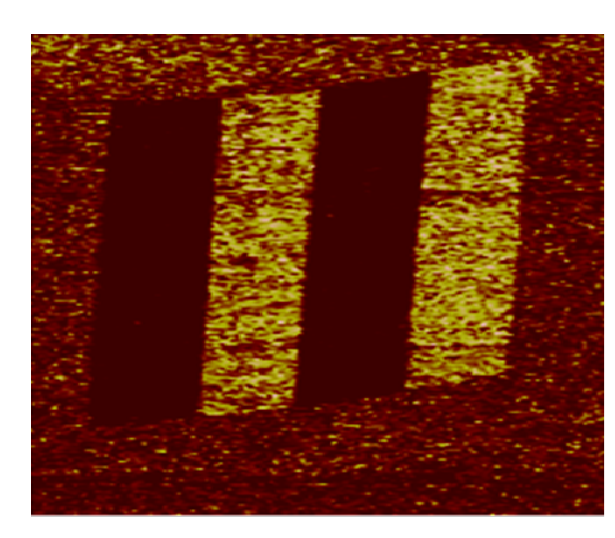}}\hfill
  \subfloat[5 nm]{\label{fig:PFM_5nm}\includegraphics[scale=0.3]{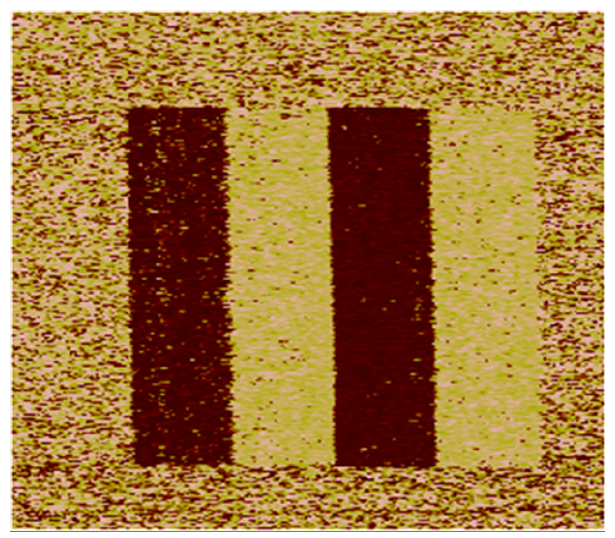}}\hfill
  \subfloat[3.6 nm]{\label{fig:PFM_3nm}\includegraphics[scale=0.3]{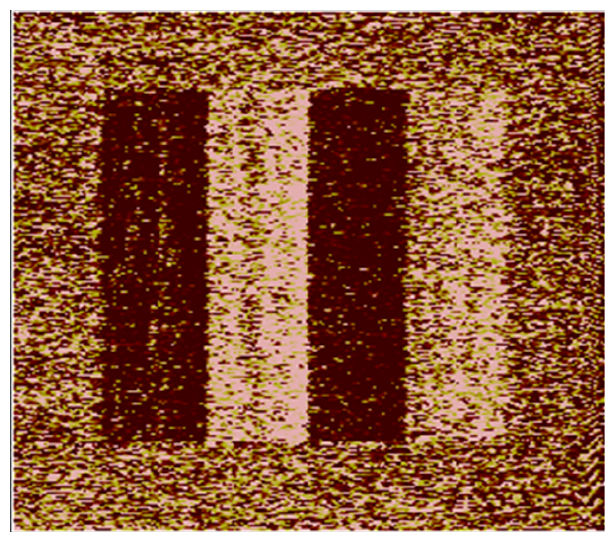}}\hfill
  \caption{PFM images for every thickness.}
  \label{fig:PFM_total}
\end{figure}

\begin{figure}[h]
  \centering
  \includegraphics[scale=0.4]{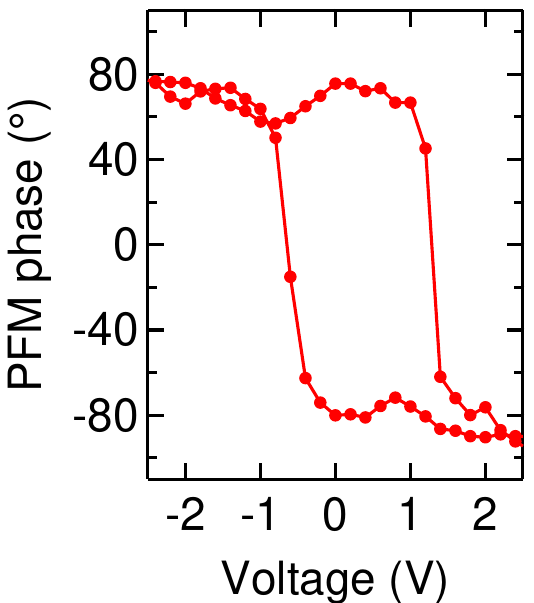}
  \caption{PFM phase loop for the 3.6~nm thin film.}
  \label{fig:PFM_phase}
\end{figure}

In addition to the 70~thin film data presented in the main manuscript, Fig.~\ref{fig:PFM_total} displays the Piezoresponse Force Microscopy (PFM) images for every thickness showing the films have been successfully poled. For the thinnest film (3.6~nm), Fig.~\ref{fig:PFM_phase} shows the PFM phase loop as a function of d.c. bias. During writing process, we applied an electric field at least twice as high as the coercive field.

\begin{figure}[h]
  \centering
  \subfloat[70 nm]{\label{fig:PEEM_70nm}\includegraphics[scale=0.27]{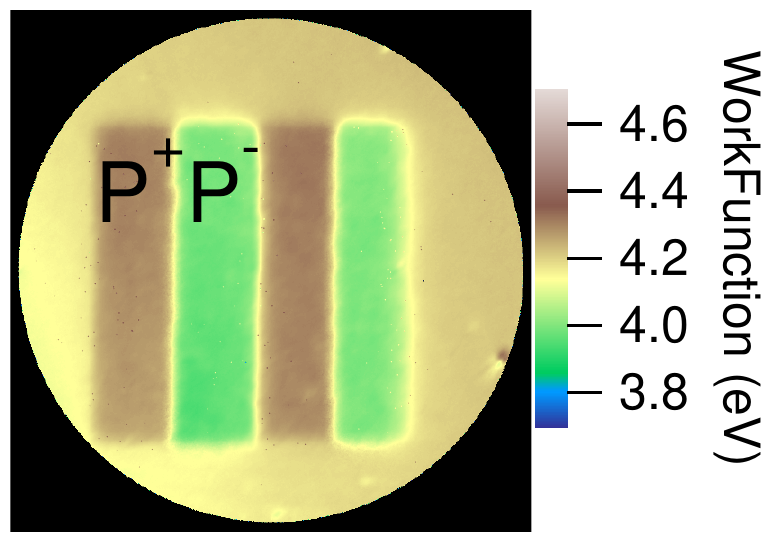}}\hfill
  \subfloat[20 nm]{\label{fig:PEEM_20nm}\includegraphics[scale=0.27]{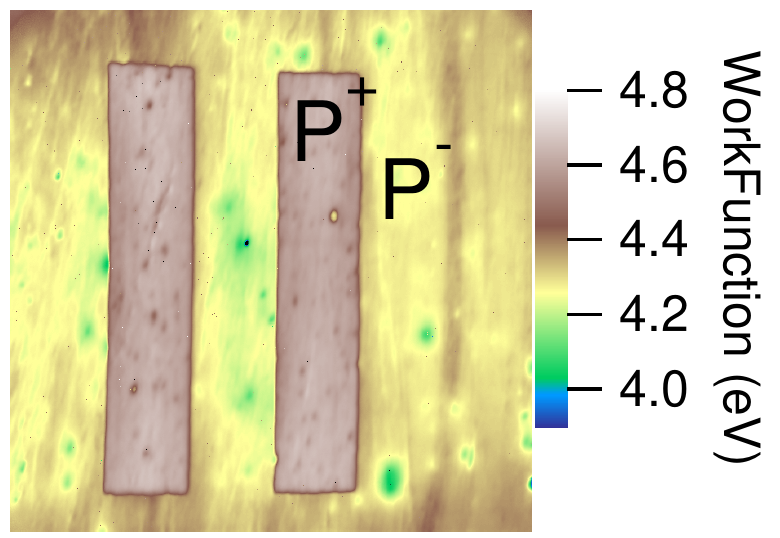}}\hfill
  \subfloat[7 nm]{\label{fig:PEEM_7nm}\includegraphics[scale=0.27]{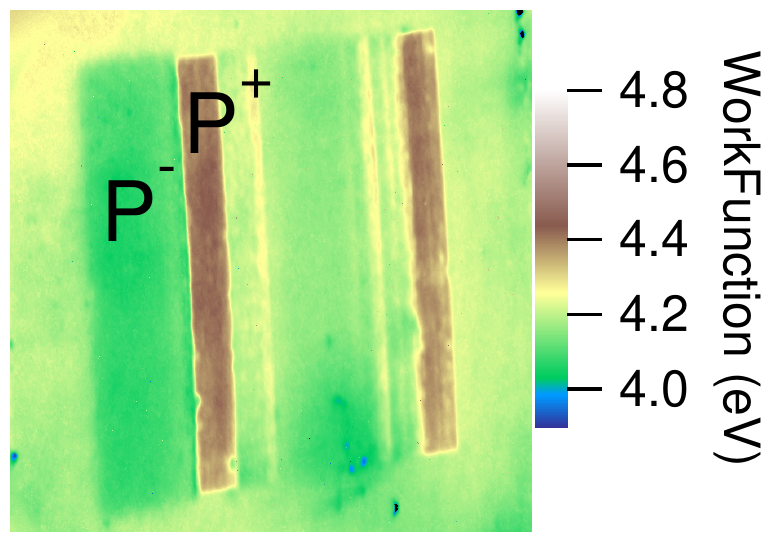}}\hfill
  \subfloat[5 nm]{\label{fig:PEEM_5nm}\includegraphics[scale=0.27]{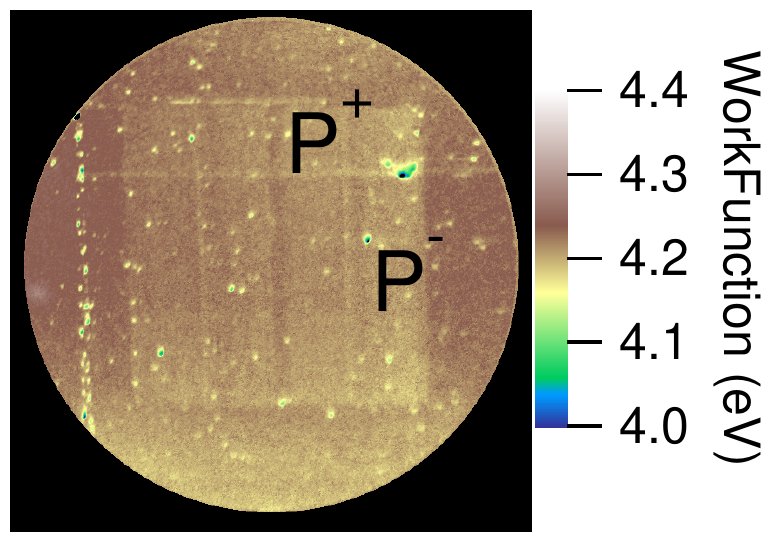}}\hfill
  \subfloat[3.6 nm]{\label{fig:PEEM_3nm}\includegraphics[scale=0.27]{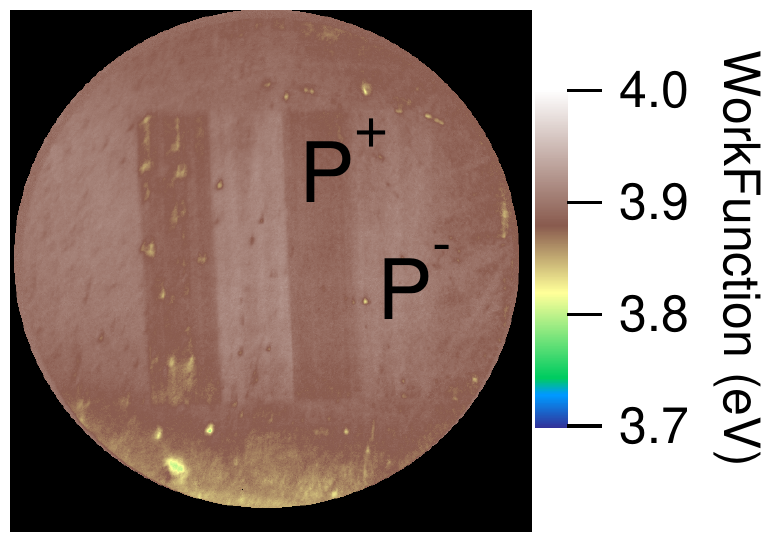}}\hfill
  \caption{Work Function maps for every thickness.}
  \label{fig:PEEM_total}
\end{figure}

\begin{figure}[h]
  \centering
  \subfloat[70 nm]{\label{fig:LEEM_70nm}\includegraphics[scale=0.27]{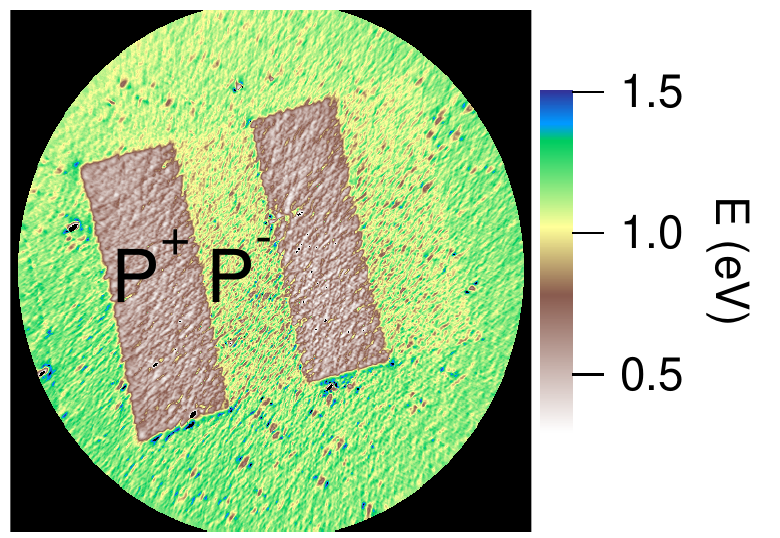}}\hfill
  \subfloat[20 nm]{\label{fig:LEEM_20nm}\includegraphics[scale=0.27]{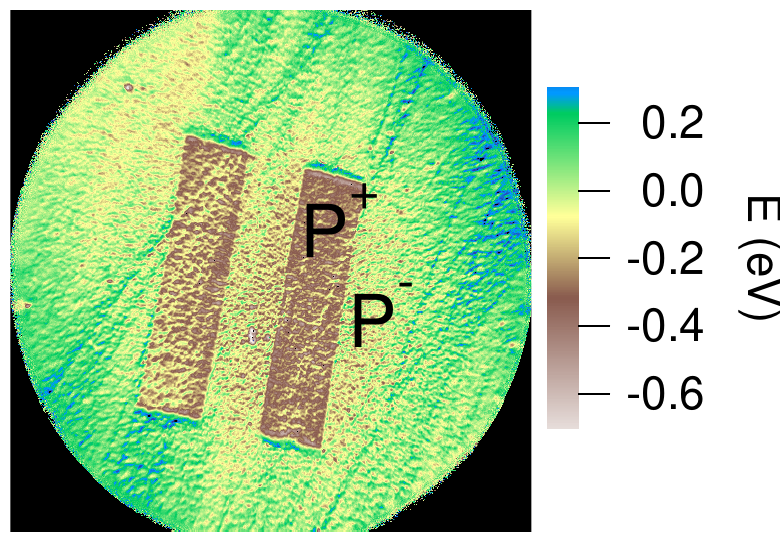}}\hfill
  \subfloat[7 nm]{\label{fig:LEEM_7nm}\includegraphics[scale=0.27]{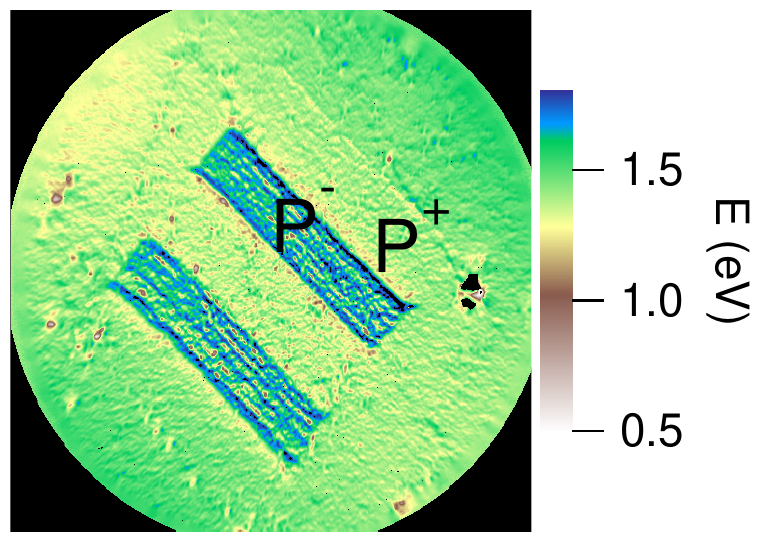}}\hfill
  \subfloat[5 nm]{\label{fig:LEEM_5nm}\includegraphics[scale=0.27]{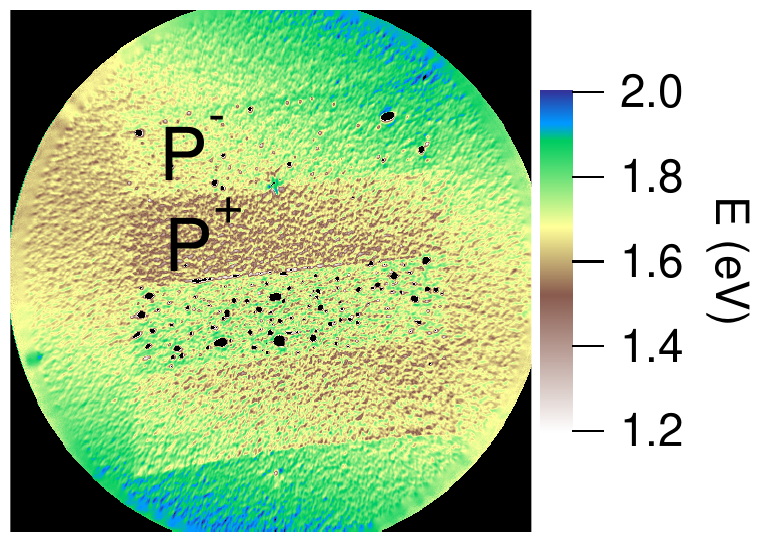}}\hfill
  \subfloat[3.6 nm]{\label{fig:LEEM_3nm}\includegraphics[scale=0.27]{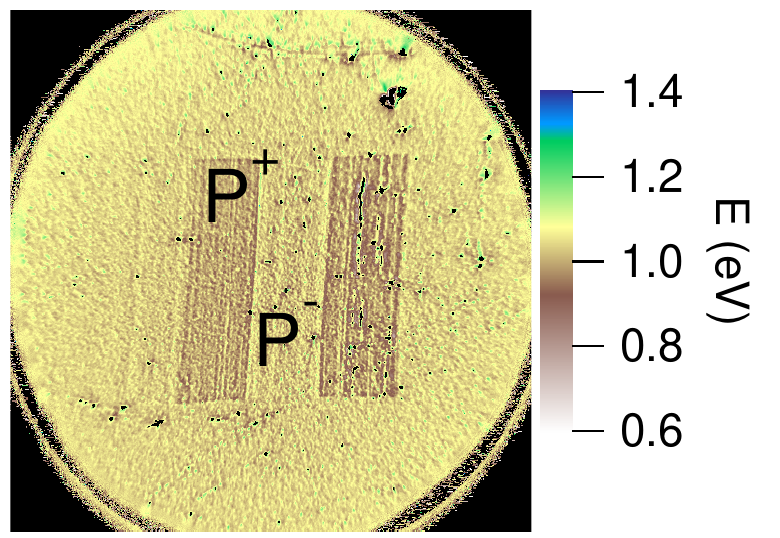}}\hfill
  \caption{MEM-LEEM transition maps for every thickness.}
  \label{fig:LEEM_total}
 \end{figure}

Fig.~\ref{fig:PEEM_total} and~\ref{fig:LEEM_total} show work function maps and MEM-LEEM transition maps for every thickness. They show clearly the different behavior between the 70, 20~nm thin films (high work function difference between P+ and P- poled regions) and the 5, 3.6~nm thin films (near-zero difference in work function). The intermediate case of 7~nm shows clear contrast (left part of Fig.~\ref{fig:PEEM_7nm} for instance). At the same time, clear contrast is visible within the P+ and P- (especially in the MEM-LEEM images) poled domains. Although the lateral resolution does not allow visualization of the nanodomains found by the numerical simulations, this additional contrast may be indirect evidence of the switching process, indicating that the formation of the stripe domains is not simultaneous across the full width of the poled domain. Oxygen vacancies can pin domain walls in, for example, PTO~\cite{he_first-principles_2003} and (Bi$_\textrm{0.85}$Pr$_\textrm{0.15}$)(Fe$_\textrm{0.95}$Mn$_\textrm{0.05}$)O$_\textrm{3}$~\cite{wen_polarization_2011}. The presence of such defects could therefore act as a nucleation center for the initial stripe domain walls. The formation of the stripe domains would then proceed outwards from the initial domain walls; indeed there is evidence in both PEEM and MEM-LEEM images for a much finer striped structure within each poled domain. A detailed study of the stripe formation is beyond the present manuscript. We believe that this illustrates single domain vs stripe domains competition around the transition thickness, likely due to slightly different boundary conditions and/or to the presence of defects in the film. It will be the subject of future work.

\subsection{Threshold Spectra Using X-ray Source}

\begin{figure}[h]
  \centering
	\includegraphics[scale=0.5]{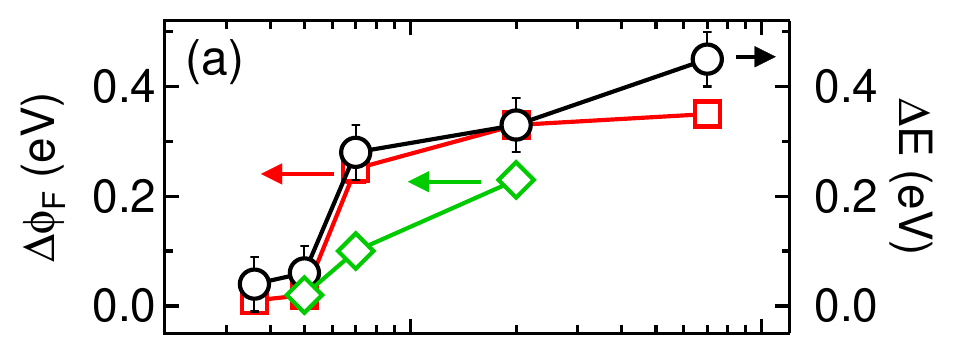}\hfill
  \caption{Thickness dependence of $\Delta\Phi_{F}$ (red squares for Hg lamp, green diamonds for X-ray source and HeI lamp) and $\Delta$E (black circles).}
  \label{fig:WF_SV_complete}
\end{figure}

Photoemission at threshold shows a cut-off energy below which electrons cannot escape the surface. It is often assumed that only secondary electrons contribute to the threshold spectra and that the complementary error function (erfc) is the correct function to deduce the work function from the rising edge of the photoemission threshold. However, the emission spectrum of the Hg light source peaks at only 4.9~eV. With such low photon energy, direct transitions might occur between p levels of the valence band and unoccupied s,d levels in the conduction band, provided of course that accessible final states lie above the vacuum level. They may give rise to intensity variations in the threshold spectra above cut-off energy and the shape of the rising edge of the photoemission threshold may be modified. In such a case the erfc parameters will no longer correctly describe threshold and inaccurate work function values may result. To check the effect of direct transitions on our work function values we took complementary image series using higher photon energy (Helium lamp h$\nu$ = 21.2~eV and X-ray source Al-K$\alpha$ h$\nu$ = 1486.70~eV) for three of the BFO films: 20, 7 and 5 nm thin films, the thicknesses around the single to stripe domains transition. The higher the photon energy, the more only true secondary electrons contribute to the photoemission threshold. Results are similar within our energy resolutions (see Fig.~\ref{fig:WF_SV_complete}) for 20 nm and 5 nm thin films. Notably, threshold widths for both types of sources are within 2\%, showing weak influence of the direct transitions on the threshold spectra. In fact, it seems that p to s,d transitions in our BFO samples leave the measured position of the low energy cut-off in the spectra largely unchanged. Therefore, the influence of direct transitions on the work function can be neglected here.

\subsection{3D generalization of Landau-Ginzburg-Devonshire to BiFeO$_{\mathrm{3}}$ Thin Films}

We start from the Ginzburg-Landau Free energy expressed in the form of the expansion with respect to the polarization P:
\begin{equation}
	F(P) = \frac{1}{2} \alpha_{\perp} \left( P_x^2 + P_y^2 \right) 
	+ \frac{1}{2} \alpha_{z} P_z^2
	+ \frac{1}{4} \beta_{1} \left( P_x^4 + P_y^4 \right)
	+ \frac{1}{4} \beta_{2} P_z^4
	+ \frac{1}{2} \beta_{3} P_x^2 P_y^2
	+ \frac{1}{2} \beta_{4} \left( P_x^2 + P_y^2 \right) P_z^2
	+ \frac{1}{6} \gamma P_z^6
	- \left( E_z + E_d \right)P_z	
\label{eq:S1}
\end{equation}

If $P_x = P_y = P_{\perp}$ then the equilibrium conditions result in the following equations:

\begin{equation}
	\alpha_z P_z + \beta_2 P_z^3 + 2 \beta_4 P_z P_{\perp}^2 + \gamma P_z^5 = E_z + E_d	
\label{eq:S2a}
\end{equation}

\begin{equation}
	\alpha_{\perp} P_{\perp} + \beta_1 P_{\perp}^3 + \beta_3 P_z^2 P_{\perp} +\beta_4 P_{\perp} P_z^2 = 0
\label{eq:S2b}
\end{equation}

Where:

\begin{equation}
	E_d = \frac{U \epsilon_0 \epsilon_d - P_z d}{\epsilon_0 \left( \epsilon_d h' + \epsilon_b d \right)} - \frac{U}{h}
\label{eq:S3}
\end{equation}

Here, $h'$ is the width of the polarized region. The total thickness of the film is $h = h' + d$, where $d$ is the dead layer width. We will assume that $d \ll h$. $\epsilon_0$ is the vacuum permittivity. $\epsilon_d$ is the dielectric constant of the dead layer and $\epsilon_b$ is the so-called background dielectric constant (which is independent of the film thickness)\cite{bratkovsky_abrupt_2000}. $U$ is the voltage between the contacts. Furthermore, $E_d$ is the depolarizing field~\cite{bratkovsky_abrupt_2000, maksymovych_ultrathin_2012}, and $E_z = U / h$.

From~(\ref{eq:S2b}), we have the choice, whether $P_{\perp} = 0$ or:

\begin{equation}
	P_{\perp}^2 = - \frac{1}{\beta_1 + \beta_3} \left( \alpha_{\perp} + \beta_4 P_z^2 \right)
\label{eq:S4}
\end{equation}

This latter equality reveals that the z-component of the polarization influences the in-plane component, and \textit{vice versa}, the magnitude of the in-plane component of the polarization influences the z-component. Now we substitute Eq.~(\ref{eq:S4}) into Eq.~(\ref{eq:S2a}) and get:

\begin{equation}
	\alpha^P P_z + \beta^P P_z^3 + \gamma P_z^5 = \frac{U \epsilon_d}{\epsilon_d h + \epsilon_b d}
\label{eq:S5}
\end{equation}

Where:

\begin{eqnarray*}
	\alpha^P & = & \alpha_z + \frac{d}{\epsilon_0 \left( \epsilon_d h' + \epsilon_b d \right)} - 2 \beta_4 \frac{\alpha_{\perp}}{\beta_1 + \beta_3}\\
			& = & \alpha^L + \frac{d}{\epsilon_0 \left( \epsilon_d h' + \epsilon_b d \right)}\\
			& \approx &  \alpha^L + \frac{d}{\epsilon_0 \epsilon_d h}\\
	\beta^p & = & \beta_2 - \frac{2 \beta_4^2}{\beta_3 + \beta_1}\\
	\alpha^L & = & \alpha_z - 2 \beta_4 \frac{\alpha_{\perp}}{\beta_1 + \beta_3}
\label{eq:S6}
\end{eqnarray*}

Notice that $\alpha^P$ is modified with respect to $\alpha_z$, and can even change sign, because of the depolarizing field and the correction due to the coupling of the in-plane component of polarization with its out-of plane component. Furthermore, $\beta^p$ is smaller compared to $\beta_2$ when all $\beta$’s are positive. This modification can even result in a negative $\beta^p$ and therefore change the second-order phase transition to a first-order one. In the case U = 0, Equation~(\ref{eq:S5}) has two stable solutions. One is $P_z = 0$, while the other is:

\begin{equation} \label{eq:S7}
P_z^2 = \frac{-\beta^p + \sqrt{\left( \beta^p \right)^2 - 4 \alpha^p \gamma}}{2 \gamma}
\end{equation}

One can easily show that such latter solution can be rewritten as:

\begin{equation} \label{eq:S11}
\frac{P_z}{P_{max}} = A \sqrt{B + \sqrt{1 - \frac{h_{eff}}{h}}}
\end{equation}

Where

\begin{eqnarray}
	D & = & \sqrt{\left( \beta^p \right)^2 - 4 \alpha^L \gamma}\\
	A & = & \frac{1}{\sqrt{B + 1}}\label{AfuncB}\\
	B & = & \frac{- \beta^p}{D}\\
	h_{eff} & = & \frac{4 d \gamma}{\epsilon_0 \epsilon_d D^2}
\label{eq:S12}
\end{eqnarray}

Equation~(\ref{eq:S11}) is the one that has been used in the manuscript to fit the data of Fig. 5b. Note that in this fitting, we allowed $A$ to take an arbitrary value, because P$_{\mathrm{max}}$ in experiment is not very well defined (one cannot consider very thick films since they become too insulating for Photoemission Microscopy). However, the resulting $A$ was numerically found to be very close to its ideal value provided in~(\ref{AfuncB}). Specifically, the ratio between the actual and ideal values for $A$ was found to be about 1.05.

Note that Equation~(\ref{eq:S11}) is, of course, valid provided that 

\begin{eqnarray*}
	1 - \frac{h_{eff}}{h} & \ge & 0\\
	B + \sqrt{1 - \frac{h_{eff}}{h}} & \ge & 0 \\
\label{eq:S13}
\end{eqnarray*}

These two conditions were met in the fit of the data of Fig. 5b for films thicker than 5.6 nm, since we numerically found that $h_{eff}$ = 5.6 nm and B = 0.16.

It is also interesting to realize that the solution of Equation~(\ref{eq:S7}) can adopt a more simple form than Equation~(\ref{eq:S11}) in some particular cases. For instance, if $\alpha^p < 0$, $\beta^p > 0$ and $\gamma = 0$ then:

\begin{equation} \label{eq:S8}
P_z^2 = \frac{- \alpha^p}{\beta^p} = P^2_{max} \left( 1 - \frac{g_{eff}}{h} \right)
\end{equation}

Where

\begin{eqnarray*}
	g_{eff} & = & \frac{d}{\epsilon_0 \epsilon_d \alpha^L}\\
	P^2_{max} & = & - \frac{\alpha^L}{\beta^p}
\label{eq:S9}
\end{eqnarray*}

Equation~(\ref{eq:S8}) has the same analytical form than the formula provided by Maksymovych \textit{et al}~\cite{maksymovych_ultrathin_2012}. However, the physical meaning of the parameters entering Equation~(\ref{eq:S8}) is different from those given in Ref.~\cite{maksymovych_ultrathin_2012}, because, here, the polarization has three non-zero Cartesian components (rather than a single one).

\end{document}